# Plasmonic photonic crystal mirror for long-lived interlayer exciton generation

*Sanghyeok Park†‡, Dongha Kim†‡, and Min-Kyo Seo†\**

†Department of Physics, Korea Advanced Institute of Science and Technology, Daejeon, 34141, Republic of Korea

‡These authors contributed equally.

*Corresponding Author: minkyo_seo@kaist.ac.kr

**ABSTRACT.** Interlayer excitons in van der Waals heterostructures of two-dimensional transition metal dichalcogenides have recently emerged as a fascinating platform for quantum many-body effects, long-range interactions, and opto-valleytronic applications. The practical implementation of such phenomena and applications requires further development of the long-lived character of interlayer excitons. Whereas material developments have successfully enhanced the nonradiative lifetime, the out-of-plane polarization nature of the interlayer excitons has made it challenging to improve the radiative lifetime with conventional photonic mirrors. Here, we propose and systematically analyze a plasmonic photonic crystal (PPhC) mirror that can increase the radiative lifetime of interlayer excitons by two orders of magnitude. Based on the vacuum field transition, the PPhC mirror supports spatially uniform radiative decay suppression over its territory, which is crucial for engineering the interlayer excitons not localized at a specific position. The PPhC mirror platform will offer new possibilities for realizing long-lived interlayer exciton-based nanodevices.

**INTRODUCTION**

The interlayer excitons in the van der Waals (vdW) heterostructure of transition metal dichalcogenide (TMDC) semiconductors have drawn significant attention because of their high binding energy[1-2] and long intrinsic lifetime[3-5], overcoming the limits of interlayer excitons of conventional III-V and II-VI semiconductor quantum wells[6-10]. Recently, state-of-the-art TMDC interlayer exciton devices have succeeded in demonstrating many-body quantum phenomena[2,11,12], macroscopic exciton transportation[13-16], and electroluminescence in atomically thin media[17]. The relatively long lifetimes of the TMDC interlayer excitons, originating from the spatial separation of the electron and valence hole planes, are critical to their success. High-quality $WSe_2$/$MoSe_2$ vdW heterostructures with hexagonal boron nitride (hBN) encapsulation, in particular, support long-lived interlayer excitons with lifetimes of several hundred nanoseconds[5], which is several orders of magnitude longer than that of the typical intralayer excitons of homogeneous TMDC layers[18-20]. The enhancement of the material-dependent non-radiative lifetime, based on the growth of high-purity crystals and the intercalation of hBN layers between the TMDC layers, has primarily been investigated to further utilize the long-lived character of the interlayer excitons[2,5,21]. Eventually, the nonradiative lifetime of interlayer excitons has recently approached the same order of magnitude as its radiative counterpart. However, practical and reliable demonstrations of high-temperature interlayer exciton condensation[2,6,11,12,22-25], the Bardeen-Cooper-Schrieffer phase[26,27], and excitonic logic devices[13-16,28-30] still require an increase in the total lifetime of the TMDC interlayer excitons.

Engineering the radiative decay rate is an alternative and convenient way to manipulate the exciton lifetime[31-33]. Whereas the nonradiative decay is controlled by the intrinsic material characteristics of the vdW layered materials used, the radiative decay can be effectively manipulated by extrinsic optical environments using a nanophotonic platform (Figure 1a). Recently, the radiative decay suppression of intralayer excitons in the TMDC mono- and multi-layer has been successfully demonstrated by a planar mirror-based platform that can minimize the local density of optical states (LDOS) of the in-plane polarization state[34-37]. Such a planar mirror platform has the advantage of preserving the material properties of the TMDC layers as well as suppressing the LDOS evenly over the two-dimensional region of interest. However, the radiative decay suppression of the interlayer excitons requires reducing the LDOS for the out-of-plane polarization, which is highly challenging due to the transverse nature of the electromagnetic waves and cannot be implemented with a conventional planar mirror.

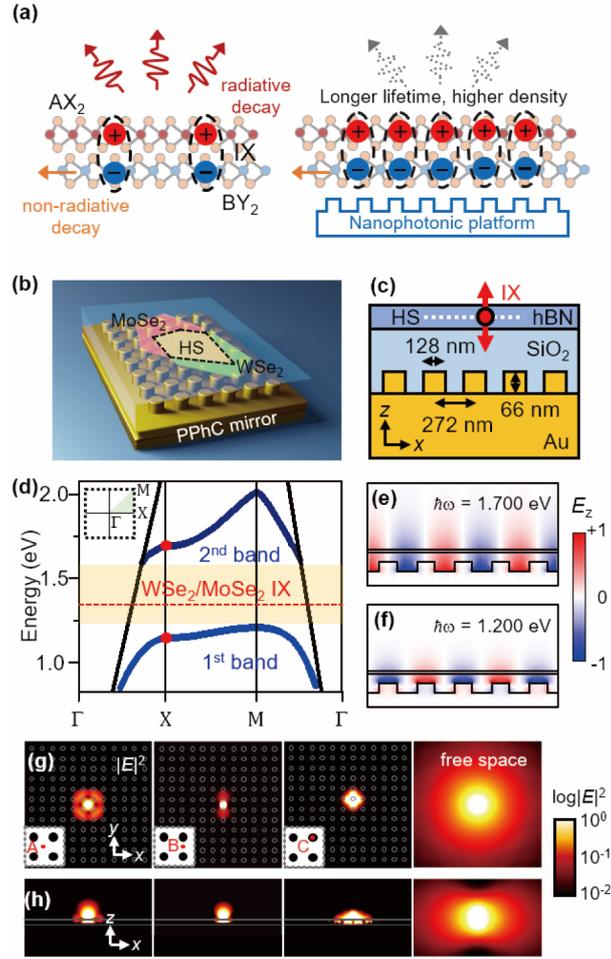

Figure 1. Nanophotonic platform for long-lived interlayer exciton generation via extreme radiative decay suppression. (a) Schematic comparison of interlayer excitons in a vdW heterostructure; (left) typical interlayer excitons in a normal vdW heterostructure and (right) long-lived interlayer excitons engineered by the nanophotonic platform. A, B atoms are transition metals (W, Mo, etc.) and X, Y atoms are chalcogens (S, Se, etc.). IX stands for interlayer exciton. (b) Illustration of the vdW heterostructure on the PPhC mirror. The green (red) plane indicates the $WSe_2$ ($MoSe_2$) monolayer. The black dashed line represents the heterostructure (HS) area. (c) Cross-sectional schematic of the employed PPhC mirror and the out-of-plane polarized emitter. The PPhC mirror consists of the 128-nm-diameter and 66-nm-height Au nanodisks in the square lattice of a period of 272 nm. The 20-nm-thick hBN layer

involving the emitter is separated from the PPhC mirror with a 130-nm-thick $SiO_2$ spacer layer. The refractive index of hBN and $SiO_2$ is 2.0 and 1.46, respectively. The white dotted line indicates the virtual vdW heterostructure layer. (d) Photonic band structure of the PPhC. The yellow shaded region indicates the photonic band gap ($\hbar\omega$ = 1.207–1.593 eV). The red dashed line indicates the photoluminescence energy level of the interlayer exciton in the $WSe_2$/$MoSe_2$ vdW heterostructure. (e, f) The $E_z$ field profile of (f) the first and (e) second band at the X point. (g, h) The electric field intensity profile of (g) the horizontal plane and (h) vertical plane. The inset shows the relative position of the dipole emitter and Au nanodisks.

This theoretical study presents a two-dimensional plasmonic photonic crystal (PPhC) mirror for generating and utilizing long-lived interlayer excitons (Figure 1b). Using the three-dimensional finite-difference time-domain method, we revealed the unique capability of the PPhC mirror to suppress the radiative decay rate of the out-of-plane polarized electric dipole emitter embedded in the planar hBN film (Figure 1c). The PPhC mirror provides a wide photonic band gap for surface plasmon polaritons, which involves the emission frequency of the interlayer exciton in the target vdW heterostructure. Although not directly attached to the planar hBN film that contains the vdW heterostructure by a dielectric spacer layer, the PPhC mirror optimally suppresses the emitter's radiation through evanescent vacuum field coupling. We demonstrated that the PPhC mirror platform reduces the Purcell factor to approximately $5.111\times10^{-3}$, corresponding to the suppression of the radiation decay rate by two orders of magnitude compared to the free space. Moreover, the PPhC mirror provides radiative decay suppression with high spatial uniformity over its unit cell and ensures reliable performance with a finite size of only approximately 1 μm, which is suitable for experimental implementation and precise spatial configuration. The transition of the vacuum field distribution to the band edge modes was used to explain the spatially uniform radiation decay suppression analytically.

**RESULTS**

In the PPhC mirror, the periodic placement of Au nanodisks on the Au film results in a wide photonic band gap in the surface plasmon polariton dispersion ranging from 1.207 to 1.593 eV, which includes the emission frequency ($\hbar\omega$ = 1.340 eV) of the interlayer exciton in the WSe$_2$/MoSe$_2$ vdW heterostructure[2-5] (Figure 1d). We note that our PPhC mirror can also cover the interlayer exciton emission frequencies of various TMDC vdW heterostructures, including the WS$_2$/MoS$_2$ (1.42 eV)[38], WSe$_2$/WS$_2$ (1.42 eV)[39], and WSe$_2$/MoS$_2$ (1.55 eV)[40] vdW heterostructures, within its large photonic band gap. Figures 1e and 1f show the cross-sectional distribution of the normal component ($E_z$) of the electric field of the first band edge mode at the M point and the second band edge mode at the X point, respectively. In the region of the photonic band gap, the LDOS for the out-of-plane polarized electric dipole emitter, which represents the interlayer exciton, is significantly suppressed. The time-averaged electric field intensity profiles in Figures 1g and 1h show the suppression of the LDOS coupled to the out-of-plane polarized dipole emitter. We examined the emitters with the frequency of 1.340 eV located at three different vertices (A, B, and C) of the PPhC unit cell. Regardless of the emitter's position, the PPhC mirror prevents the emitted field from propagating in any direction in three dimensions. The strongly concentrated evanescent profile of the field indicates that the emitter has few optical states to couple its energy into radiation. A comparison with the field profile of the emitter in free space clarifies this point. Employed permittivity of Au were extracted from experimental reference[41].

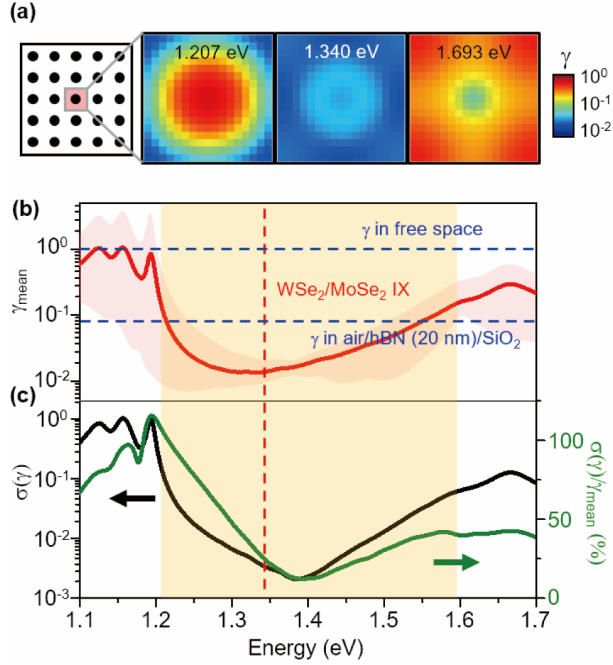

Figure 2. Purcell factor suppression for the out-of-plane polarized emitter. (a) Profile of the Purcell factor of the out-of-plane polarized electric dipole emitter depending on the position over the single unit cell (the red shaded box) located at the center of the 5×5 PPhC array mirror. The normalized radiative decay rate to the free space corresponds to the Purcell factor ($\gamma$). We plot the Purcell factor suppression profiles at three different frequencies of the first band edge (M point), the center of the band gap, and the second band edge (X point): $\hbar\omega$ = 1.207, 1.340, and 1.693 eV. (b) Spectra of the averaged Purcell factor $\gamma_{mean}$ (the solid red line) and the range from the minimum to maximum values (the red shaded area). The yellow shaded region and the red dashed line indicate the photonic band gap and the photoluminescence energy of the interlayer exciton in the $WSe_2/MoSe_2$ heterostructure, respectively. (c) Spectra of the standard deviation of the Purcell factor, $\sigma(\gamma)$, and the coefficient of variation, $\sigma(\gamma)/\gamma_{mean}$.

The TMDC interlayer excitons are not localized at a specific position and diffuse over micrometer-scale distances[5,13,14]. Thus, long-lived interlayer exciton generation requires the effective suppression of the LDOS over the two-dimensional region of interest. Figure 2a shows a two-dimensional plot of the radiative decay rate suppression depending on the emitter position over a unit cell located at the center of the 5×5 PPhC array mirror. We calculated the Purcell factor ($\gamma$), the radiative decay rate normalized to that in free space. The suppression of the Purcell factor at the emission frequency of the $WSe_2/MoSe_2$ interlayer exciton (1.340 eV) ranges from $2.067\times10^{-2}$ to $9.111\times10^{-3}$, which corresponds to the radiative lifetime enhancement from 48 to 110 times. The photonic band gap of the SPPs enables the PPhC mirror to efficiently and uniformly suppress the radiative coupling of the electric dipole emitter of the out-of-plane polarization over a two-dimensional space. On the contrary, the suppression of the Purcell factor loses its spatial uniformity at the frequencies of the band edges. At the M-point of the 1st band (1.207 eV), the Purcell factor varied from $8.956\times10^{-3}$ to $5.608\times10^{-1}$. At the X-point of the 2nd band (1.693 eV), the Purcell factor varied from $3.900\times10^{-2}$ to $4.548\times10^{-1}$. It is worth noting that the spatial distribution of the Purcell factor is correlated to that of the normal component ($E_z$) of the electric field at the band edges.

To examine the performance of the PPhC mirror, we employed the mean and standard deviation ($\gamma_{mean}$ and $\sigma(\gamma)$) of the Purcell factor depending on the emitter position. Figure 2b shows $\gamma_{mean}$ (the solid red line) and the band between the maximum and minimum values (red shaded area) of the Purcell factor suppression as a function of the frequency of the emitter. At the target emission frequency of the interlayer exciton (1.340 eV), the PPhC mirror suppresses the Purcell factor down to $1.369\times10^{-2}$ on average when compared to free space, increasing the radiative lifetime by approximately 73 times. Sandwiched between the low-refractive-index media (air and $SiO_2$) and because of its refractive index of 2.0[42], the hBN layer assists the radiative decay suppression to some extent[43,44] (blue dotted line in Figure 2b). The Purcell factor is inversely proportional to the cubic of the surrounding medium's refractive index[31], and the original dipole emitter and the image dipoles induced in the low-refractive-index media destructively interfere with each other. The effect of the thickness of the hBN layer is discussed in detail later, together with the results in Figure 5a. The PPhC mirror efficiently suppressed both vertical and horizontal radiation and supported the radiative lifetime enhancement of more than 50 times over a broad spectral range from 1.258 to 1.408 eV. The standard deviation of the Purcell factor and the coefficient of variation $\sigma(\gamma)/\gamma_{mean}$ quantify the spatial uniformity of the suppressed LDOS (Figure 2c). The coefficient of variation was minimized to approximately 12% at a frequency of 1.393 eV. At the interlayer exciton emission frequency of 1.340 eV, $\sigma(\gamma)$ and $\sigma(\gamma)/\gamma_{mean}$ are approximately $3.559 \times 10^{-3}$ and 26.00%, respectively.

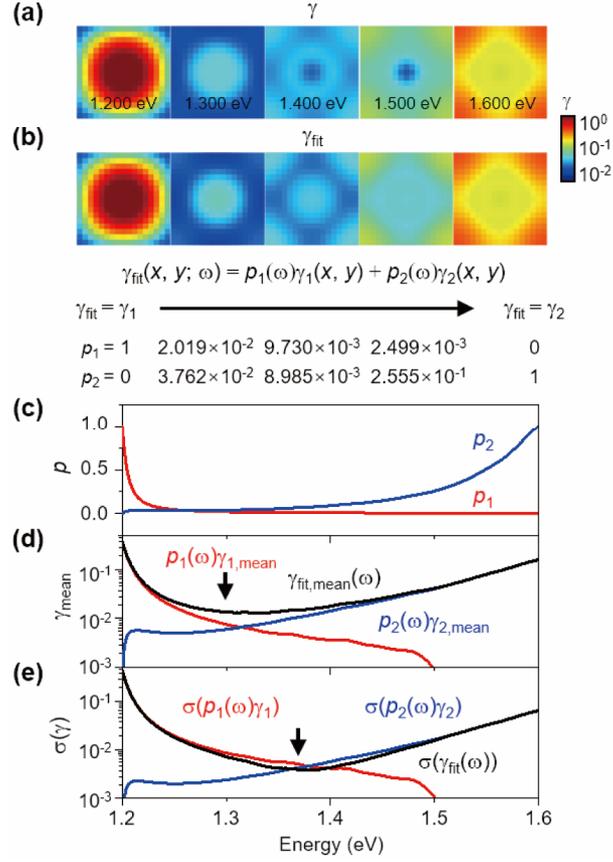

Figure 3. Transition of the vacuum field distribution across photonic band gap. (a) Simulated Purcell factor map ($\gamma(x, y; \omega)$) from 1.200 to 1.600 eV. (b) Fitted Purcell factor map ($\gamma_{fit}(x, y; \omega)$) from 1.200 to 1.600 eV. Fitting of $\gamma_{fit}(x, y; \omega)$ was conducted by the provided formula. Weighting parameter $p_1$ ($p_2$) gradually changes from 1 to 0 (0 to 1) in photonic band gap. (c) Spectra of weighting parameter $p_1$ (red) and $p_2$ (blue). (d) Contribution of the first ($p_1(\omega)\gamma_{1,mean}$, red) and the second band ($p_2(\omega)\gamma_{2,mean}$, blue) in average Purcell factor ($\gamma_{fit,mean}$, black) during the transition of vacuum field distribution. (e) Contribution of the first ($\sigma(p_1(\omega)\gamma_1)$, red) and the second band ($\sigma(p_2(\omega)\gamma_2)$, blue) in standard deviation of the Purcell factor ($\sigma(\gamma)$, black) during the transition of vacuum field distribution. Black arrows in (d, e) indicate the spectral location of minimum $\gamma_{fit,mean}$ and $\sigma(\gamma_{fit})$, respectively.

We revealed that the spatially uniform radiation decay suppression of the PPhC mirror originates from the smooth transition of the vacuum field distribution between the patterns at the first and second band edges. The field distribution at the first band edge (the M point) is spatially complementary to that at the second band edge (the X point), and the first and second band edge modes concentrate their electric fields in the region of the Au nanodisk (at the center of the unit cell) and the region out of the Au nanodisk (at the corners of the unit cell), respectively. It is well known that such a difference in field concentration results in a photonic band gap[45]. Figure 3a shows the calculated distribution of the Purcell factor, $\gamma(x, y; \omega)$, over a unit cell depending on the frequency across the band gap. This study analyzed $\gamma(x, y; \omega)$ with a two-dimensional fitting employing the conical combination, $\gamma_{fit}(x, y; \omega) = p_1(\omega)\gamma_1(x, y) + p_2(\omega)\gamma_2(x, y)$, where $\gamma_1(x, y)$ and $\gamma_2(x, y)$ are the numerically calculated distributions of the Purcell factor at the frequencies of the first and second band edges, respectively. We find the set of $p_1$ and $p_2$ minimizing the multidimensional squared Euclidean distance between $\gamma_{fit}(x, y; \omega)$ and $\gamma(x, y; \omega)$. Figure 3b shows the fitted distribution of the Purcell factor for the selected frequencies and confirms the smooth transition of the vacuum field distribution. The centered hole observed in $\gamma(x, y; \omega)$ at the frequencies of 1.400 eV and 1.500 eV does not appear in $\gamma_{fit}(x, y; \omega)$ because the conical fitting neglects the possibility of destructive interference between the vacuum field distributions originating from the first and second bands.

As shown in Figure 3c, the non-negative, conical weighting parameter $p_1$ ($p_2$) continuously varies from one to zero (0 to 1) as the frequency changes from 1.200 to 1.600 eV. It is worth noting that $p_1$ rapidly decreases after leaving the first band edge, whereas $p_2$ increases gradually on the way to the second band edge. In the photonic band gap, the dispersion relation lies in the space of the real-valued frequency and the imaginary-valued wavenumber. The reciprocal slope of the evanescent dispersion ($d\text{Im}(k)/d\omega$) is less stiff around the second band edge than around the first band edge, which explains the asymmetric behavior of the conical weighting parameters depending on the frequency[45]. The mean and standard deviation of the Purcell factor was predicted using weighting parameters (Figure 3d and 3e). The mean of the Purcell factor follows the conical combination as $\gamma_{\text{fit,mean}}(\omega) = p_1(\omega)\gamma_{1,\text{mean}} + p_2(\omega)\gamma_{2,\text{mean}}$, and, as shown in Figure 3d, has a minimum of $1.309\times10^{-2}$ at a frequency of 1.332 eV, which matches with the result in Figure 2b. The combined standard deviation of the Purcell factor is determined by

$$\sigma(\gamma_{\text{fit}}(\omega)) = \sqrt{\sigma^2(p_1(\omega)\gamma_1) + \sigma^2(p_2(\omega)\gamma_2) + 2\text{cov}(p_1(\omega)\gamma_1, p_2(\omega)\gamma_2)},$$

where cov($a$, $b$) is the covariance between $a$ and $b$. Because of the complementary relation between $p_1(\omega)\gamma_1(x, y)$ and $p_2(\omega)\gamma_2(x, y)$, cov($p_1(\omega)\gamma_1, p_2(\omega)\gamma_2$) is always negative. The smaller the elements, $\sigma(p_1(\omega)\gamma_1)$ and $\sigma(p_2(\omega)\gamma_2)$, and the more equal they are, the smaller the combined standard deviation. In the proposed PPhC mirror, the condition for minimizing the combined standard deviation holds around the center of the photonic band gap. The predicted frequency and value of the minimized combined standard deviation were 1.378 eV and $3.771\times10^{-3}$, respectively (Figure 3e), which is consistent with the numerically calculated results (Figure 2c).

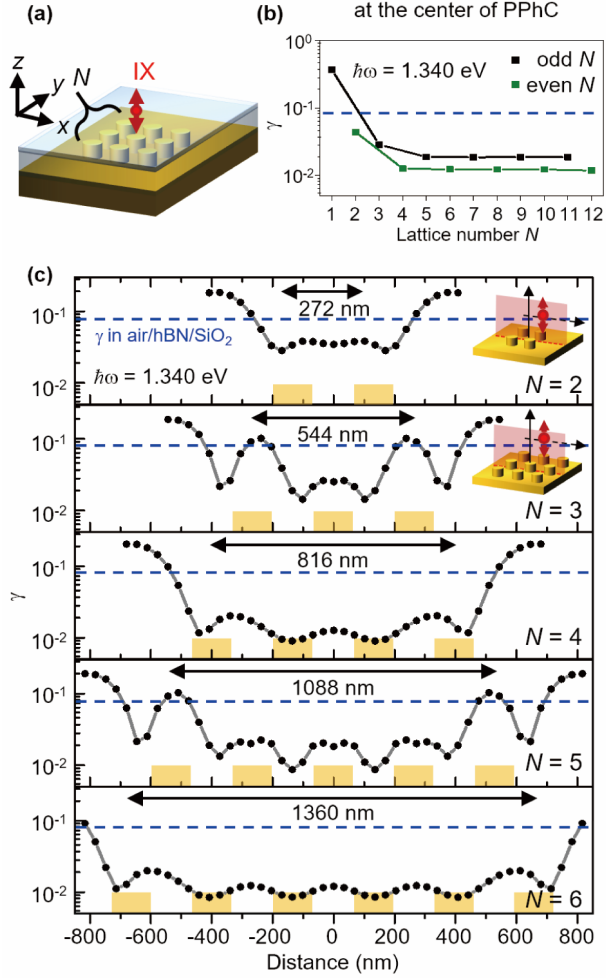

Figure 4. Operation of the finite-sized PPhC mirror. (a) Schematic of the finite-sized PPhC mirror of an $N \times N$ square lattice. The red arrow indicates the out-of-plane polarized electric dipole emitter. (b) The Purcell factor suppression depending on the size of the PPhC mirror. The emitter is located at the center of the finite-sized PPhC mirror where the Au nanodisk is concentric (farthest) for odd (even) $N$. The black and green dots indicate the Purcell factor when $N$ is odd and even, respectively. The blue dashed line indicates the Purcell factor of the emitter positioned in the middle of the 20-nm-thick hBN layer sandwiched between the air and the $SiO_2$ spacer without the PPhC mirror. (c) Scanning of the Purcell factor as a function of the emitter position across the PPhC mirrors with different sizes from $N = 2$ to $N = 6$. The yellow

boxes indicate the positions of the Au nanodisks projected to the plane of scanning. The insets schematize the plane of scanning (shaded red) for $N = 2$ and $N = 3$.

The PPhC mirror guarantees reliable performance even with a 1 μm order size. The radiative decay suppression of the finite-sized PPhC mirrors with an $N \times N$ square lattice of Au nanodisks was calculated (Figure 4a). Figure 4b shows the Purcell factor suppression for an emitter of 1.340 eV depending on the size of the PPhC mirror. Located at the center of the PPhC mirror, the emitter is aligned to (for odd $N$) or farthest from (for even $N$) the Au nanodisk; the suppressed Purcell factor oscillates according to the parity of $N$. The radiative decay suppression is stronger at the midpoint between the Au nanodisks than at the nanodisk center, which matches the result in Figure 2a. From $N = 4$, the finite-sized PPhC mirror supports the radiative decay suppression reaching that of the infinite array.

Figure 4c shows the radiative decay suppression depending on the position of the emitter for the PPhC mirrors with different sizes from $N = 2$ to $N = 6$. The Purcell factor along a line passing through the center of the PPhC mirror was examined (see the inset schematics in Figure 4c). It should be noted that for $N \geq 3$, almost the same level of radiative decay suppression is obtained uniformly in all unit cell regions with the exception of the outermost unit cells. For example, the PPhC mirror with $N = 4$ ensures sufficient suppression of radiative decay over the region of three unit cells corresponding to a side length of 816 nm. As a result, the PPhC mirror requires only a few unit cells to suppress the Purcell factor of the electric dipole emitter with an out-of-plane polarization state. Thus, it is expected that the PPhC mirror will be a useful platform for the spatial engineering of the LDOS in a two-dimensional space with a 1 μm scale resolution.

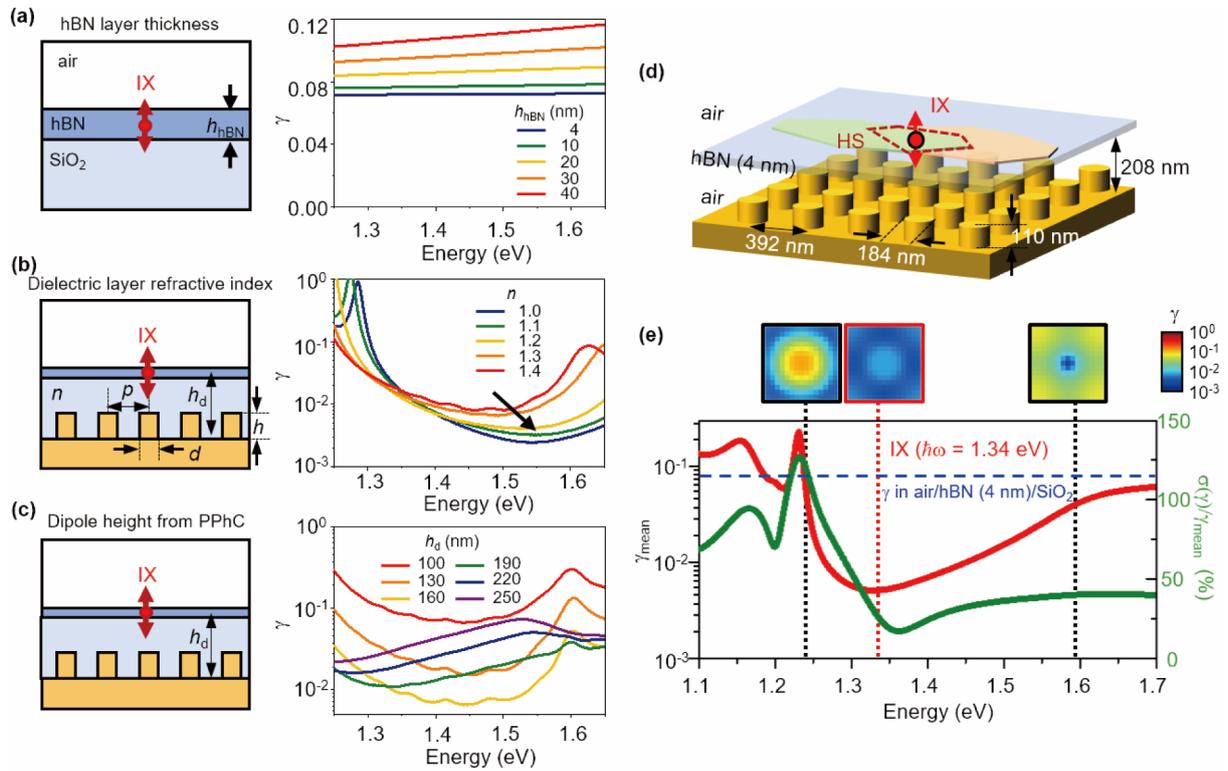

Figure 5. Optimization of the PPhC mirror platform. (a) Effect of the thickness of the hBN layer on the Purcell factor. The emitter is located in the middle of the hBN layer sandwiched between the air superstrate and the $SiO_2$ substrate without the PPhC mirror. (b) Effect of the refractive index of the spacer layer on the Purcell factor. The refractive index (*n*) of the spacer layer changes from 1.0 to 1.4; however, the structural parameters are scaled as changing the refractive index of the spacer layer. (c) Effect of the thickness of the dielectric spacer layer ($h_d$) on the Purcell factor. (d) Schematic of the optimized PPhC mirror: the diameter and height of the Au nanodisk, 184 and 110 nm, the period of the square lattice, 392 nm, and the thickness of the air spacer, 208 nm. The vdW heterostructure is located in the middle of the 4-nm-thick hBN layer. (e) Spectra of the averaged Purcell factor, $\gamma_{mean}$, (the solid black line) and the coefficient of variation, $\sigma(\gamma)/\gamma_{mean}$, (the solid green line). The statistics was examined over the unit cell located at the center of the 5×5 PPhC array mirror. The Purcell factor profiles over the central unit cell for the three representative frequencies of the first and second band edges (1.245 eV, 1.593 eV) and the WSe$_2$/MoSe$_2$ interlayer exciton emission are plotted.

We investigated the effect of the structural parameters of the PPhC mirror platform systemically to optimize radiative decay suppression. First, the effect of the thickness of the hBN layer on the Purcell factor of the out-of-plane polarized emitter was addressed (Figure 5a). As the thickness decreases from 40 to 4 nm, the Purcell factor decreases from $1.055 \times 10^{-1}$ to $7.179 \times 10^{-2}$ at a frequency of 1.340 eV. As aforementioned, the hBN layer sandwiched between the low-refractive-index substrate and superstrate contributes to suppressing the radiative decay of the internal out-of-plane polarized electric dipole emitter[43,44]. In the planar interface between two different dielectric media, the electric dipole in the higher-refractive-index medium develops a bound polarization distribution whose effect is equivalent to the out-of-phase image dipole in the lower-refractive-index medium. For the out-of-plane polarized electric dipole emitter, the image dipole emitters result in destructive interference, suppressing radiation in the horizontal direction. As the thickness of the hBN layer is reduced and the dipole emitter moves closer to the interfaces, the destructive interference-based radiative decay suppression becomes stronger with convergence. Second, the effect of the refractive index ($n$) of the dielectric spacer layer beneath the hBN layer was examined (Figure 5b). The structural parameters of the PPhC was scaled (Au nanodisk diameter $d$, height $h$, periodicity $p$, and the spacer thickness $h_d$) as linear as the refractive index changes from 1.4 to 1.0; $nd$ = 186.88 nm, $np$ = 397.12 nm, $nh$ = 96.36 nm, and $nh_d$ = 204.4 nm. The thickness of the hBN layer was fixed at 20 nm, and the emitter was located at the center of the 5×5 PPhC array mirror. As the refractive index of the spacer layer decreases, the frequency that minimizes the Purcell factor is blue-shifted, and the minimum value of the Purcell factor is gradually lowered. Third, the radiative decay suppression was calculated depending on the spacing distance ($h_d$) of the emitter from the PPhC mirror (Figure 5c). The SiO$_2$ spacer is beneath the 20-nm-thick hBN layer. For an emission frequency of 1.340 eV, the Purcell factor was optimized down to 9.627 ×10$^{-3}$ at a spacing distance of 160 nm. At spacing distances longer and shorter than the optimized condition, the weaker coupling to the PPhC mirror and the stronger field absorption in the Au medium, discourages radiative decay suppression.

Based on the aforementioned methods, we developed the optimized PPhC mirror-based platform for long-lived interlayer exciton generation, as shown in Figure 5d. The platform consists of a free-standing 4-nm-thick hBN layer and a 5×5 PPhC array mirror. The spacing distance between the hBN layer and PPhC mirror was 208 nm. Figure 5e shows the performance of the optimized radiative decay suppression platform. At the interlayer exciton emission frequency, the Purcell factor averaged over the spatial area of the center unit cell is suppressed down to 5.111×10$^{-3}$, which corresponds to a radiative lifetime enhancement of 195 (24.38) times compared to the free space (the homogeneous hBN medium). The coefficient of variation was as small as 22.96%. Considering that the radiative and nonradiative lifetimes of interlayer excitons in the vdW heterostructure of high-quality TMDCs are of similar order[5], such strong suppression of the Purcell factor can almost double the total lifetime. The saturation density and diffusion length of excitons are proportional to the total lifetime and its square root[2,13-16], respectively. Thus, we believe that the effective suppression of the radiative decay rate would lead to actual increase in the condensation temperature and transport range of interlayer excitons.

**CONCLUSION**

We theoretically present the two-dimensional PPhC mirror that can suppress the radiative decay rate of interlayer excitons of the out-of-plane polarization nature in a vdW heterostructure two orders of magnitude lower than the free space emission. It is analytically revealed that the smooth transition of the vacuum field distribution passing the photonic band gap enables the PPhC mirror to support the Purcell factor suppression with high spatial uniformity over the two-dimensional region of interest. Systematic optimization of the PPhC mirror platform demonstrates not only radiative decay suppression breaking the stereotyped limits of conventional planar mirror systems[34-36] but also the possibility of engineering the LDOS in a two-dimensional space as desired with a one-micrometer-scale resolution. It is expected that the PPhC mirror will provide a promising platform for the lifetime extension of interlayer excitons for fundamental studies and applications of high-temperature exciton condensation[2,6,11,12,22-25] and long-distance interaction[26,28] and transport[13,14,28-30,46,47]. We also believe that the PPhC mirror will boost the realization of novel excitonic devices, such as excitonic memory cells[48,49] and low-loss optoelectronic transistors[13-16,28-30].

# AUTHOR INFORMATION

**Corresponding Author**

Min-Kyo Seo – Department of Physics, Korea Advanced Institute of Science and Technology, Daejeon, Daehak-Ro 291, Republic of Korea; Email: minkyo_seo@kaist.ac.kr

**Authors**

Sanghyeok Park – Department of Physics, Korea Advanced Institute of Science and Technology, Daejeon, Daehak-Ro 291, Republic of Korea

Dongha Kim – Department of Physics, Korea Advanced Institute of Science and Technology, Daejeon, Daehak-Ro 291, Republic of Korea

**Author Contributions**

‡S.P and D.K. contributed equally to this work.

**Notes**

The authors declare no competing financial interest.

# ACKNOWLEDGMENT

M.-K.S. acknowledges support by KAIST Cross-Generation Collaborative Lab project and the National Research Foundation of Korea (NRF) (2020R1A2C2014685, 2020R1A4A2002828). D. K. acknowledges support by acknowledges support by NRF (2015H1A2A1033753).

**Table of Contents (TOC)**

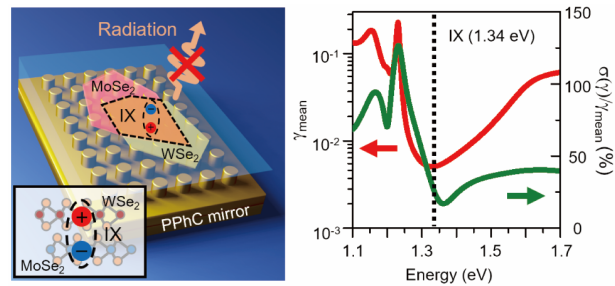